 \documentstyle[prb,aps,multicol,epsfig]{revtex}
\begin{document}
\draft \date{\today} \title
 {Theory of Type-II Superconductors with Finite London Penetration Depth}
\author{Ernst Helmut Brandt}
\address{Max-Planck-Institut f\"ur Metallforschung,
   D-70506 Stuttgart, Germany}
\maketitle

\begin{abstract}

  Previous continuum theory of type-II superconductors of various
shapes with and without vortex pinning in
an applied magnetic field and with transport current,
is generalized to account for a finite London penetration
depth $\lambda$. This extension is particularly important
at low inductions $B$, where the transition to the Meissner
state is now described correctly, and for films with thickness
comparable to or smaller than $\lambda$. The finite width of
the surface layer with screening currents and the correct
dc and ac responses in various geometries follow naturally
from an equation of motion for the current density in which
the integral kernel now accounts for finite $\lambda$.
New geometries considered here are thick and thin strips
with applied current, and ``washers'', i.e.\ thin film squares
with a slot and central hole as used for SQUIDs.
\end{abstract}
\pacs{PACS numbers: \bf 74.60.Ec, 74.60.Ge, 74.55.+h}
    \begin{multicols}{2}
    \narrowtext

\section{Introduction}  % 1

  The statics and dynamics of the Abrikosov vortex lines \cite{1}
and of the two dimensional (2D) pancake vortices in layered
superconductors \cite{2} recently was formulated within continuum
approximation in terms of an equation of motion of the
current density ${\bf j}$ inside the superconductor.\cite{3}
The resulting integral equation for the space and time dependent
${\bf j}(x,y,z,t)$ contains the assumed {\it constitutive law} for
the electric field ${\bf E}_v$ caused by the motion of vortices. In
general, the law ${\bf E = E}_v({\bf j, B})$ depends on the
current density ${\bf j}$ and on the local induction ${\bf B}$ and
describes, e.g., free flux flow, pinning, and thermally  activated
depinning of vortices.

  Besides this, the equation of motion for ${\bf j}$ depends on the
{\it geometry} of the given problem. Various geometries have been
considered: Thin long strips,\cite{4} thin circular disks and
rings,\cite{5} thin rectangular platelets or films,\cite{6} thick
strips,\cite{7} and thick circular disks or short cylinders,\cite{8}
always in a magnetic field applied perpendicular to the plane of the
superconductor. Further numerical solutions for the statics
of thin film superconductors of various shapes within the
Bean model are given by Prigozhin.\cite{9}
In all these geometries, the current density ${\bf j}$ or the
sheet current in films
${\bf J}(x,y,t) =(J_x, J_y) = \int {\bf j}(x,y,z,t) \,dz$
have only one component (flowing along the strip or circulating
in the disk) or may be derived from a scalar potential $g(x,y,t)$
according to $J_x =\partial g / \partial y$,
$J_y =-\partial g / \partial x$  since $\nabla\cdot {\bf J}=0$.
The equation of motion describes thus a scalar ($j$ or $g$) which
depends on only one or two spatial coordinates.

   A more general continuum formulation including also
transport currents applied by contacts, applying in principle
to arbitrary three dimensional (3D) geometry, and accounting also
for a finite lower critical field $H_{c1}$ and for a general
reversible thermodynamic field $H(B)$, is presented in
Ref.~\onlinecite{10}, where it is applied to the problem of
the geometric barrier.\cite{11} The elegance of the formulation
\cite{4,5,6,7,8,10} in terms of an equation of motion for
the current density is that it explicitly accounts for the
applied magnetic field or transport current, which in other
formulations enter via a separate boundary condition for a
differential equation. Our integral equation automatically accounts
for the inhomogeneous magnetic field outside the superconductor,
without any need to calculate this infinitely extended field
explicitly during the time integration since all spatial integrations
extend only over the finite specimen cross section. The applied
field and applied current enter via the screening currents that
flow on the specimen surface, the information about the
geometry is contained in a pre-calculated integral kernel,
and the properties of the superconductor enter via one single
constitutive law $E = E_v(j,B)$ if $B=\mu_0 H$ may be
assumed,  or via two laws if some nontrivial law
$B(H) \ne \mu_0 H$ is used.

  So far, this continuum description \cite{4,5,6,7,8,10}
considered the limit where both the vortex spacing $a$ and the
London penetration depth $\lambda$ are assumed to be much smaller
than all other relevant lengths. There are, however, two
situations where a finite London depth $\lambda$ should be
considered:

(i) In bulk superconductors, at small inductions
$B$ the Meissner state should follow in the limit $B\to 0$.
So far, with the above assumption $E = E_v (j,B)$ one has
$E = 0$ for $B\to 0$ if the free flux flow law
or similar laws $E_v \propto B$ are used. In real type-II
superconductors, however, one may have $E \ne 0$ in a
surface layer of thickness $\lambda$ if $j$ varies with time,
since one has $E = \mu_0\lambda^2 \partial j/\partial t$
in the Meissner-London limit (this London equation means that
the electric field accelerates the Cooper pairs).

(ii) In thin films with thickness $d$, both cases $d > \lambda$
and $d< \lambda$ should be described correctly by a continuum theory.
In particular, for $d\ll \lambda$ the effective penetration depth
of thin films \cite{2,12} $\Lambda = \lambda^2 / d$ may become
macroscopically large and comparable with the film width.

  In the present paper our previous continuum electrodynamics of
type-II superconductors \cite{4,5,6,7,8,10} is generalized to
allow for an arbitrary
London penetration depth $\lambda$. The resulting integral
equations for $j$ in various geometries correctly describe
all limiting cases and are easily solved numerically. With finite
$\lambda$ the stability and speed of the numerics is even
improved since this $\lambda$ provides a well defined inner cutoff
length, whereas with $\lambda = 0$ the inner cutoff depends on
the chosen grid spacing.

  In the following sections, equations of motion for the current
density in type-II superconductors with finite London depth $\lambda$
are derived for various geometries. For illustration, some selected
numerical results are presented for each geometry, namely,
current and magnetic field profiles, magnetization curves, ac
susceptibilities, and the current stream lines in thin films.

\section{Voltage--Current Law}  % 2

   In type-II superconductors within continuum approximation the
local electric field $E$ is composed of two parts, originating
from the vortex motion and from the Meissner surface currents,
  \begin{eqnarray}  % 1
  {\bf E = E}_v + {\bf E}_M \,.
  \end{eqnarray}

  The first term ${\bf E}_v({\bf j, B})$ in general is a nonlinear
function of the current density ${\bf j}$ due to pinning and
thermal or quantum depinning (creep). It may be anisotropic
(a) if the Hall effect of vortex motion is accounted for, or
(b) if the superconductor is anisotropic, or
(c) if ${\bf j}$ is not perpendicular to ${\bf B}$.
This latter case is still little understood but should
occur in most really three dimensional geometries, e.g., in a
superconductor cube with applied magnetic field; one possible
model description in this case uses two critical current
densities $j_{c\perp}$ and $j_{c\|}$ for the current components
perpendicular and parallel to ${\bf B}$. Though the parallel
current density $j_\|$ does not exert a Lorentz force on the
vortices it nevertheless can trigger a helical instability
\cite{13,14} and vortex cutting, \cite{15} which leads to a
finite $j_{c\|}$. Furthermore, as shown by Gurevich\cite{16}
some combinations of nonlinearity and anisotropy in
voltage--current models may lead to unstable current distributions.
  In the following geometry examples, for transparency I shall consider
isotropic ${\bf E}_v$ since the extension to anisotropic ${\bf E}_v$
is straightforward. Isotropic voltage--current laws have the form
  \begin{eqnarray}  % 2
  {\bf E}_v({\bf j},B) = \rho_v(j,B) \, {\bf j} \,,
  \end{eqnarray}
where $\rho_v(j,B)$ is the resistivity caused by vortex motion,
and $j = |{\bf j}|$, $B = |{\bf B}|$, with ${\bf j \perp B}$.
In the free flux flow limit (zero pinning or high current density)
one has $\rho_v = \rho_{\rm ff} \approx (B/B_{c2}) \rho_n$
where $B_{c2}$ is the upper critical field and $\rho_n$ the normal
resistivity just above $B_{c2}$. For thermally activated depinning
many experiments and theoretical models yield a logarithmic
activation energy $U(j) = U_0 \ln(j_c /j)$,
which gives $E_v = E_c\exp(-U/kT) = E_c (j/j_c)^n$ with creep
exponent $n = U_0/kT \gg 1$ and (arbitrary)
``threshold criterion'' $E_c = E(j_c)$. With Eq.~(2) this means
$\rho_v = \rho_c (j/j_c)^\sigma$ where $\sigma = n-1 \gg 1$.
A simple model which combines this creep model and the free
flux flow limit reads \cite{10}
  \begin{eqnarray}  % 3
  \rho_v(j,B) = {\rho_n \over B_{c2}} B {(j/j_c)^\sigma
     \over 1 + (j/j_c)^\sigma } \,.
  \end{eqnarray}

  The second term in Eq.~(1) is caused by the current flowing in
the Meissner state and is given by the dynamic London equation,
  \begin{eqnarray}  % 4
   {\bf E}_M = \mu_0 \lambda^2 {\partial {\bf j} \over \partial t} \,.
  \end{eqnarray}
This term describes the acceleration of the massive charge
carriers (Cooper pairs) by an electric field,
$\partial j / \partial t \propto E$. Thus, a finite London depth
$\lambda$ in principle may be introduced into our continuum
description \cite{4,5,6,7,8,10} by using the full voltage--current
law,
  \begin{eqnarray}  % 5
   {\bf E} = {\bf E}_v({\bf j, B}) + \mu_0 \lambda^2
     {\partial {\bf j} \over \partial t}  \,.
  \end{eqnarray}

 This replacement works well in the case of {\it linear} response,
where it yields a frequency ($\omega$) dependent complex resistivity
  \begin{eqnarray}  % 6
   \rho(\omega) = \rho_{\rm ff}  + i\omega\mu_0 \lambda^2
  \end{eqnarray}
in the absence of vortex pinning. In this case the linear integral
equation for $j$ reduces to an eigen value equation from which the
complex ac susceptibility of the superconductor for a given geometry
follows as an explicit sum.\cite{17}

  When $E_v(j,B)$ is {\it nonlinear}, the integral equation for
$j({\bf r},t)$ has to be solved by numerical integration over time.
It turns out that this time integration becomes unstable when the
modified voltage--current law (5) is inserted. However, when the
term containing $\lambda^2$ is incorporated into the integral
kernel as shown in Sct.~3, then the numerics remains stable and
even becomes more stable and faster than it was with $\lambda=0$.

\section{Thick strips with finite $\lambda$}   % 3

   To fix ideas we first derive the modified integral equation for
the geometry of an infinite strip or bar with rectangular
cross section filling the volume $-a \le x \le a$, $-b \le y \le b$,
$-L/2 \le z \le L/2$, $L \gg a,b$, see Fig.~1.
In the general 3D geometry, with
${\bf r}=(x,y,z)$, the vector potential ${\bf A_j(r)}$ caused by
the current density ${\bf j(r)} = -\mu_0^{-1} \nabla^2 {\bf A_j}$
in the gauge $\nabla\cdot {\bf A} = 0$ is given by
  \begin{eqnarray}  % 7
  {\bf A_j(r)} = \mu_0 \int\! d^3 r' {{\bf j(r}')
    \over 4\pi |{\bf r-r}'| } \,,
  \end{eqnarray}
with the integral taken over the volume where the current flows.
Indeed, using $\nabla^2 |{\bf r-r}'|^{-1} =-4\pi\delta_3({\bf r-r}')$
($\delta_3$ is the 3D delta function) one verifies that Eq.~(7)
yields $\nabla^2 {\bf A_j(r)} = -\mu_0 {\bf j(r)}$. For a long
strip or bar, both ${\bf j} = j(x,y)\, {\bf\hat z}$ and
${\bf A} = A(x,y)\, {\bf\hat z}$ are directed along $z$.
Integrating Eq.~(7) over $z$ and writing from now on
${\bf r} = (x,y)$ we obtain
  \begin{eqnarray}  % 8
  A_j({\bf r}) = \mu_0 \int\! d^2r'j({\bf r}')\,
      Q_{\rm bar}({\bf r,r}')
  \end{eqnarray}
with the 2D integral kernel
  \begin{eqnarray}  % 9
  Q_{\rm bar}({\bf r,r}') = {1\over 2\pi} {\rm asinh}{L/2 \over
  |{\bf r-r}'|} \approx {1\over 2\pi} \ln {L \over |{\bf r-r}'|} \,.
  \end{eqnarray}
The integral (8) is over the rectangular strip cross section,
$-a\le x\le a$ and $-b \le y \le b$, but actually Eq.~(8) applies
to strips with cross sections of any shape. If the current
distribution is symmetric the integration (8) may be restricted to
one quarter of the rectangular cross section, e.g.,
  \begin{eqnarray}  % 10
  A_j(x,y) = \mu_0 \!\int_0^a \!\!\! dx'\! \int_0^b \!\!\! dy'
  j(x', y')\, Q_{\rm bar}^{\rm sym} (x,y;x',y') \,.
  \end{eqnarray}
For strips with a magnetic field $H_a$ applied along
$y$, one has $j(x,y) = -j(-x,y)= j(x,-y) = -j(-x,-y)$, thus
  \begin{eqnarray}  % 11
  Q_{\rm bar}^{\rm sym}=Q_{\rm bar}^H =  {1\over 4\pi}
  \ln{ (x_+^2+y_-^2)(x_+^2+y_+^2) \over (x_-^2+y_-^2)(x_-^2+y_+^2)}
  \end{eqnarray}
with $x_\pm =x\pm x'$, $y_\pm =y\pm y'$. For strips with transport
current $I_a$ along $z$ one has $j(x,y) =j(-x,y) =j(x,-y) =j(-x,-y)$,
thus
  \begin{eqnarray}  % 12
  Q_{\rm bar}^{\rm sym}=Q_{\rm bar}^I = ~~~~~~~~~~~~~~~~~~~~~~~~~~~
  ~~~~~~~~~~~~~~ \nonumber \\  {1\over 4\pi} \ln{ L^8 \over
  (x_-^2+y_-^2)(x_-^2+y_+^2)(x_+^2+y_-^2)(x_+^2+y_+^2) } \,.
  \end{eqnarray}
Note that the strip length $L$ has dropped out in Eq.~(11) but not
in Eq.~(12). Therefore, some electrodynamic properties of long
strips with applied current (e.g.\ their self induction) {\it depend
logarithmically on the strip length} $L$, while for strips in a
magnetic field usually the limit $L \to \infty$ may be taken. Strips
with oblique applied field ${\bf H}_a = (H_{ax}, H_{ay})$, or with
both applied $H_a$ and $I_a$, have a lower symmetry
and the integration (8) then has to be taken over the half or full
cross section of the strip rather than over a quarter.

  To obtain an explicit dynamic equation for the current density
$j(x,y,t)$ one has to incorporate the applied vector potential
${\bf A}_a$, which for the strip may be chosen along $z$,
${\bf A}_a = A_a(x,y,t)\, {\bf \hat z}$. In general,
$A_a(x,y,t)$ may have two parts originating from an applied
perpendicular magnetic field or induction
${\bf B}_a(t) = \mu_0{\bf H}_a(t) = (B_{ax}, B_{ay})$
and from an applied electric field ${\bf E}_a=E_a(t)\,{\bf\hat z}$
which drives the transport current $I_a$,
  \begin{eqnarray}  % 13
  A_a(x,y,t) = A_a^B + A_a^E \,, \nonumber \\
  {\bf B}_a = \nabla \times (A_a^B {\bf\hat z}),~~~
  E_a = -\dot{A}_a^E \,,
  \end{eqnarray}
where the dot denotes the time derivative. For example, with
$B_a \| y$ one has $A_a^B = -x B_a$, and with $E_a(t=0)=0$
one has $A_a^E = - \int_0^t E_a(t') dt'$. Writing the total vector
potential as $A = A_j + A_a$ we obtain from Eq.~(8),
  \begin{eqnarray}  % 14
  \mu_0 \!\int\! d^2r' j({\bf r}',t) Q_{\rm bar}({\bf r,r}')
   =A_j({\bf r},t)=A-A_a\,.
  \end{eqnarray}
For the strip the induction law $\dot{\bf B} =-\nabla\times{\bf E}$
yields $E=-\dot{A}$. Inserting $E$ from Eq.~(5) we have for strips
  \begin{eqnarray}  % 15
  \dot{A} = -E_v(j,B) - \mu_0 \lambda^2 \partial j /\partial t\,,
  \end{eqnarray}
thus the r.h.s. of Eq.~(14) may be written as
  \begin{eqnarray}  % 16
  A_j({\bf r},t) =-\!\!\int \!E_v(j,B)\,dt -\mu_0\lambda^2
   j({\bf r},t)  -\! A_a({\bf r},t) \,.
  \end{eqnarray}
Inserting this in Eq.~(14) and shifting the term
$\mu_0 \lambda^2 j({\bf r},t)$ to the left under the integral
we obtain
  \begin{eqnarray}  % 17
  \mu_0\! \int\! d^2r' j({\bf r}',t) [\, Q_{\rm bar}({\bf r,r}')
     + \lambda^2 \delta_2({\bf r-r}') \,] \nonumber \\
     = - \int\! E_v({\bf r},t)\, dt - A_a({\bf r},t) \,,
  \end{eqnarray}
where $\delta_2({\bf r}) = \delta(x) \delta(y)$ is the 2D delta
function. Taking the time derivative of Eq.~(17) and introducing
the inverse integral kernel \cite{18,19}
  \begin{eqnarray}  % 18
  K({\bf r,r}')  = [\, Q_{\rm bar}({\bf r,r}') +
              \lambda^2 \delta_2({\bf r-r}') \,]^{-1} \,.
  \end{eqnarray}
we arrive at the equation of motion for $j({\bf r},t)$:
  \begin{eqnarray}  % 19
  {\partial j({\bf r},t) \over \partial t} = - \mu_0^{-1} \!\!
  \int\!\! d^2 r' K({\bf r,r}') [E_v(j,B) +\dot{A}_a({\bf r}',t) ]
  \end{eqnarray}
with $\dot{A}_a({\bf r}',t) = -x'\dot B_{ay}(t) +y'\dot B_{ax}(t)
 -E_a(t)$ and with $Q_{\rm bar}$ from Eqs.~(9), (11), or (12).
In Eq.~(19) only the electric field caused by the vortex motion
$E_v$ enters explicitly. The London length $\lambda$ enters the
inverse kernel $K({\bf r,r}')$ defined by
  \begin{eqnarray}  % 20
  \int\! d^2 r'' K({\bf r,r}'') [\, Q_{\rm bar}({\bf r}'',{\bf r}') +
   \lambda^2 \delta_2({\bf r}''-{\bf r}') \,] \nonumber \\
   = \delta_2({\bf r-r}')\,.
  \end{eqnarray}

  The kernel $K({\bf r,r}')$ may be computed by a
matrix inversion as follows. First, a spatial  grid
${\bf r}_i = (x_i, y_i)$ is chosen with appropriate weights $w_i$
such that the integrals over the strip cross section are well
approximated by a sum,
  \begin{eqnarray}  % 21
  \int\! d^2 r f({\bf r}) \approx \sum_i f({\bf r}_i) w_i \,.
  \end{eqnarray}
Then we express the definition (20) by such a sum,
  \begin{eqnarray}  % 22
  \sum_i K_{li} \,(\, Q_{ij}^{\rm bar} w_i + \lambda^2
    \delta_{ij} \,)  = \delta_{lj} \,,
  \end{eqnarray}
where $Q_{ij} = Q({\bf r}_i,{\bf r}_j)$ and $\delta_{ij}$
equals 1 if $i=j$ and 0 else. Solving Eq.~(22) for
the matrix $K_{ij}$  one finds
  \begin{eqnarray}  % 23
  K_{ij} =(\,Q_{ij}^{\rm bar}w_i +\lambda^2 \delta_{ij} \,)^{-1} .
  \end{eqnarray}
The accuracy of this method is considerably increased by choosing
a nonequidistant grid with narrow spacing near the specimen
surface and by taking appropriate diagonal terms $Q_{ii}^{\rm bar}$
as described in the Appendix of Ref.~8a. From Eq.~(23) one sees that
finite $\lambda^2$ increases the (positive) diagonal terms of the
matrix to be inverted; this makes the matrix inversion more stable.

  For numerics we need the equation of motion  (19) for
${\bf j(r},t)$ in discrete form,
  \begin{eqnarray}  % 24
  {\partial j_i \over \partial t} = - \mu_0^{-1} \! \sum_j
   K_{ij} \, [\, E_v(j_j,B_j) +\dot{A}_{aj} \,] \,,
  \end{eqnarray}
where the vectors $j_i(t) = j({\bf r}_i,t)$,
$B_i(t) =B({\bf r}_i,t)$, and
$\dot A_{ai}(t) =\dot A_a({\bf r}_i,t) =
 -x_i \dot B_{ay}(t) +y_i \dot B_{ax}(t) -E_a(t)$ are functions
of the time $t$. The matrix $K_{ij}$, Eq.~(23), is independent of
time and has to be computed {\it only once} for a given geometry
and given $\lambda$. Equation (24) is easily integrated over
time $t$ starting with $j_i(t=0) = 0$ and then switching on
the applied fields $B_a$ and/or $E_a$. The resulting magnetic
moment per unit length ${\bf m}(t) = (m_x, m_y)$  and total
transport current $I_a(t)$ (along $z$) are then obtained as
integrals over the strip cross section,
  \begin{eqnarray}  % 25, 26
  {\bf m}(t) &=& \int_{-a}^a \!\! dx\! \int_{-b}^b \!\! dy
  (-{\bf\hat x}y +{\bf\hat y}x) \,j(x,y,t) \nonumber \\
  &\approx& \sum_i (-{\bf\hat x}y_i+{\bf\hat y}x_i)\,j_i(t)\,w_i
                \,,\\
  I_a(t) &=& \int_{-a}^a \!\! dx\! \int_{-b}^b \!\! dy \,
  j(x,y,t) \approx \sum_i j_i(t)\,w_i \,.
  \end{eqnarray}
Note that the contribution to ${\bf m}(t)$ of the U-turn of
the currents at the strip ends (integrals over the $x$ and
$y$ components of ${\bf j}$, amounting to exactly
${1\over2}{\bf m}$) is already considered in Eq.~(25).
Note further that, though in experiments usually the applied
current $I_a(t)$ is imposed, the theory  considers
a spatially constant electric field $E_a(t)$ along $z$ which
drives the current $I_a(t)$ that results from the calculation.

  Figures 2--6 show the current density $j(x,y)$ and the magnetic
field lines of ${\bf B}$ (i.e.\ the contour lines of $A$) for strips
with aspect ratio $b/a = 0.4$ in various cases, cf.\ Fig.~1.
In Fig.~2  the strip is in the Meissner state, i.e., no vortices
have penetrated. Shown is the screening current density for finite
London depth $\lambda= 0.025 a$ for a strip in applied field (top)
and for a strip with applied current (bottom). Similar figures for
strips with quadratic cross section are depicted in
Ref.\ \onlinecite{19}. Note the sharp peak of $j(x,y)$ in the four
corners, which has finite height. This high local current density
favors the nucleation of vortex quarter loops from the corners of
the strip when the applied magnetic field or current exceed a
certain threshold.

  Figures 3 and 4 show current density and field lines for strips
in an increasing applied field $H_a$ for two values of the London
depth $\lambda/a= 0.025$ and 0.1. Since the assumed voltage--current
law is very steep, $E_v \propto j^n$ with $n=101$,
a flat saturation of $j(x,y)$ occurs at the critical
value $j_c$, like in the Bean model. The field of full penetration
of this strip is \cite{7}
$H_p = (j_c b/\pi)[(2a/b)\arctan(b/a) +\ln(1+a^2/b^2)] = 0.4945 aj_c$
for $b/a =0.4$ and $H_p =2.52\cdot(2bj_c)$ for $b/a=0.001$.

  Figures 5 and 6 show the same strips but with increasing applied
current $I_a$ and with no field applied. In this case, when the critical
current $I_c =4abj_c$ of the strip is reached, vortex rings penetrate
continuously and annihilate in the center of the strip. With
further increasing $I_a > I_c$, the dissipation and voltage drop
increase steeply. More results for strips with both
applied field and current will be published elsewhere.

  Figures 7 and 8 show magnetization loops $m(H_a)$ of strips in
perpendicular applied field $H_a(t) = H_0 \sin\omega t$ at two
amplitudes $H_0$, for five values of the London depth $\lambda$, and
for two creep exponents $n$ entering in $E_v(j)=E_c (j/j_c)^n$.
Figure 7 is for a thick strip ($b/a=0.4$) and Fig.~8 for a thin strip
($b/a=0.001$). With our dimensionless units $a=j_c=E_c=\mu_0=1$,
the used circular frequency $\omega =1$ corresponds
to $\omega = E_c / (\mu_0 j_c a^2)$ in physical units.
Note that with increasing  $\lambda$ the hysteresis loop becomes
more narrow and finally collapses to a line, i.e., the magnetic
response becomes {\it reversible}. This is so since with
increasing $\lambda$ the screening current density decreases
and can no longer depin vortices, except when $H_a$ is large
or $n$ is small.

\section{Thin Strips}   % 4

   This section considers the limit of thin strips, $b\ll a$,
with $B_a$ applied perpendicular (along $y$) and $I_a$ applied
parallel (along $z$) to the strip. In this limit only the current
density integrated over the film thickness matters, called
{\it sheet current} and directed along $z$ (like $j$),
  \begin{eqnarray}  % 27
  J(x,t) = \int_{-b}^b \! j(x,y,t)\, dy \,.
  \end{eqnarray}
Integrating Eq.~(17) over $y$ we obtain
(see also Ref.~\onlinecite{20})
  \begin{eqnarray}  % 28
  \mu_0\! \int_{-a}^a \!\! dx' J(x',t)\, \Big[\, {1\over 2\pi}
  \ln{L \over |x-x'|} +\Lambda \delta(x-x') \,\Big] \nonumber \\
     = - \int\! E_v(x,t)\, dt - A_a(x,t) \,,
  \end{eqnarray}
where $\Lambda = \lambda^2 /d$ is the effective penetration
depth of thin films with thickness $d = 2b < \lambda$.
Here I have used the approximate constancy of the current
density along $y$, $j(x,y) \approx J(x)/d$. Initially it
was not clear to me if thin film expressions of the type (28)
describe also the {\it dynamics} of superconductor strips and not
only the {\it statics}, which was successfully considered, e.g.
in the static Bean model calculations of thin disks, \cite{21}
thin strips, \cite{22,23,24,25} and ellipses. \cite{26} But
recently we have proven \cite{27} that not only $j({\bf r},t)$
but also the electric field $E({\bf r},t)$ is practically
constant over the film thickness; deviations from this
constancy occur only near the (penetrating or exiting)
flux front but are restricted to a transverse length scale of
order $d$. This result was obtained in the limit
$\lambda \to 0$, but it applies all the more for finite $\lambda$.
Thus, the dynamics of the sheet current $J \approx jd$
is well described by Eq.~(28).

  Inverting Eq.~(28) and taking the time derivative we obtain the
equation of motion for the sheet current in thin strips,
  \begin{eqnarray}  % 29
  \mu_0 \dot J(x,t) = -\! \int_{-a}^a \!\!\! dx' K(x,x')
  [\,E_v(x',t) +\dot{A}_a(x',t) \,]
  \end{eqnarray}
with the inverse integral kernel
  \begin{eqnarray}  % 30
  K(x,x') = \Big[\, {1\over 2\pi} \ln{L \over |x-x'| }
  +\Lambda \delta(x-x') \,\Big]^{-1}
  \end{eqnarray}
and with $E_v(x',t) = E_v(j,B)$ depending on $j=J(x',t)/d$ and
$B = |B(x',t)|$, and with $\dot A_a(x',t) =-x'\dot B_a(t) -E_a(t)$.
For strips with either applied field $B_a$ or applied current $I_a$,
the integration in Eq.~(29) may be restricted to half the strip width,
$0\le x \le a$. For $B_a \ne 0$, $I_a=0$ one has $J(x)=-J(-x)$
and a symmetric kernel [cf.\ Eq.~(11)]
  \begin{eqnarray}  % 31
  K_B(x,x') = \Big[\, {1\over 2\pi} \ln{x+x' \over |x-x'| }
  +\Lambda \delta(x-x') \,\Big]^{-1} \,.
  \end{eqnarray}
For $B_a =0$, $I_a \ne 0$ one has $J(x)=J(-x)$ and a symmetric
kernel [cf.\ Eq.~(12)]
  \begin{eqnarray}  % 32
  K_I(x,x') = \Big[\, {1\over 2\pi} \ln{L^2 \over |x\! - \!x'|
  (x\! +\! x')} + \Lambda \delta(x\! -\! x') \Big]^{-1} .
  \end{eqnarray}
Note that the strip length $L$ has dropped out in the kernel (31)
but not in the kernel (32). The complex resistivity of thin strips
thus in general depends on the logarithm of the strip length.

  Figures 9 and 10 show the profiles of the sheet current $J(x)$
and perpendicular induction $B_y(x)$ of thin strips with increasing
applied magnetic field (Fig.~9) or current (Fig.~10) for various
values of the effective penetration depth $\Lambda = \lambda^2 /d$;
$d=2b$ is the strip thickness and $2a$ the strip width,
here $b/a=0.001$. Note that for larger $\Lambda$ the profiles
$J(x)$ become smoother, but the bend of $J(x)$ at the point where
$J$ starts to deviate from the critical sheet current $J_c=dj_c$
remains sharp. For $\Lambda/a > 0.2$, the sections of $J(x)$ where
$|J| < J_c$  are almost straight lines in Fig.~9 and almost
parabolas in Fig.~10.

  Figure 11 shows  $J(x)$ and $B_y(x)$ in thin strips which are
exposed to a high magnetic field $H_a \gg H_p$ before a current
$I_a$ is applied, ranging from 0 to the critical current
$I_c = 4abj_c$. First, when $I_a=0$, one has
$J(x) \approx J_c {\rm  sign}(x)$, and finally, when
$I_a\approx I_c$, one has $J(x) \approx J_c$ (equal signs would
apply in the Bean limit $n\to \infty$). At intermediate
currents $I_a < I_c$, in the half strip  $0< x <a$ one has
$J \approx J_c =$ constant, and in the other half $-a <x <0$
the profile $J(x)$ is symmetric about the point $x=-a/2$. For
$\Lambda=0$, $n \gg 1$, the known Bean profiles for thin strips
with transport current apply to this half strip. \cite{22,23,24,25}

\section{Axial symmetry}   % 5

   The above method for incorporating finite $\lambda$ can also be
applied to axially symmetric problems, i.e.\ to superconductors
with an axis of rotational symmetry in a magnetic field applied
parallel to this axis (along $y$). A simple example are disks or
short cylinders with radius $a$ and thickness $d=2b$, but also other
shapes like rings, toruses, cones, ellipsoids, etc.\  are easily
described by introducing a $y$ dependent radius $a(y)$. In all
these cases the current density and vector potential have only one
component directed along the azimuthal unit vector ${\bf\hat
\varphi}$, ${\bf j} = j(r,y) {\bf\hat \varphi}$ and ${\bf A} =
A(r,y) {\bf\hat \varphi}$, with $r=(x^2 +z^2)^{1/2}$. Integrating
the 3D Eq.~(7) over the angle $\varphi=\arctan(z/x)$ we obtain the
vector potential $A_j$ caused by the current density $j$
circulating in an axially symmetric conductor,
  \begin{eqnarray}  % 33
  A_j(r,y) = \mu_0 \!\int_{-b}^b \!\!\! dy'\! \int_0^{a(y')}
  \!\!\!\!  dr' j(r',y')\, Q_{\rm ax} (r,y;r',y')
  \end{eqnarray}
with the kernel \cite{8}
  \begin{eqnarray}  % 34, 35
  Q_{\rm ax}(r,y;r',y') = f(r,r',y-y') \,, \\
  f(r,r',\eta) = \! \int_0^\pi \!\! {d\varphi \over 2\pi}
  {-r'\cos\varphi \over (\eta^2+r^2+r'^2 -2rr'\cos\varphi)^{1/2}} \,.
  \end{eqnarray}
When the conductor has a symmetry plane at $y=0$, then the integration
over $y'$ in Eq.~(33) may be restricted to $0 \le y\le b$ if the
kernel $Q_{\rm ax}$ is replaced by a symmetric kernel,
  \begin{eqnarray}  % 36
  Q_{\rm cyl}(r,y;r',y') = f(r,r',y-y') +f(r,r',y+y') \,.
  \end{eqnarray}
If in addition the radius is constant (like in disks and cylinders),
Eq.~(33) simplifies to
  \begin{eqnarray}  % 37
  A_j(r,y) = \mu_0 \!\int_0^b \!\!\! dy'\! \int_0^a \!\!\!
  dr' j(r',y')\, Q_{\rm cyl} (r,y;r',y') \,.
  \end{eqnarray}
Comparing Eqs.~(33) and (37) with Eqs.~(8) and (10) for the strip
or bar, one easily verifies that the analogue to Eq.~(17) for
axial symmetry is identical to Eq.~(17) but with
$Q_{\rm bar}({\bf r,r}')$ replaced by $Q_{\rm ax}({\bf r,r}')$,
Eq.~(34) or $Q_{\rm cyl}({\bf r,r}')$, Eq.~(36), where now
${\bf r}=(r,y)$, ${\bf r'}=(r',y')$. In axial symmetric problems
no transport current and thus no electric field is applied but
only a magnetic field or induction
 ${\bf B}_a= \nabla \times (A_a(r,y,t) {\bf\hat\varphi}) $
 $ =(B_{ar}, B_{ay})$. In general this applied field may be
inhomogeneous, e.g.\ when levitation forces are to be computed. If
${\bf B}_a$ is homogeneous one has $A_a(r,y,t) =-{r\over2} B_{a}(t)$.
With this notation, for the general axially symmetric geometry
the equation of motion for the circulating current density is
identical to Eq.~(19) but with a different inverse kernel
  \begin{eqnarray}  % 38
  K({\bf r,r}')  = [\, Q_{\rm ax}({\bf r,r}') +
              \lambda^2 \delta_2({\bf r-r}') \,]^{-1} \,,
  \end{eqnarray}
with $Q_{\rm ax}$ from Eq.~(34). If the superconductor has a
symmetry plane at $y=0$ and the applied field is homogeneous
(or symmetric about the plane $y=0$) one may replace
 $Q_{\rm ax}$ in Eq.~(38) by  $Q_{\rm cyl}$ from Eq.~(36),
  \begin{eqnarray}  % 39
  K({\bf r,r}')  = [\, Q_{\rm cyl}({\bf r,r}') +
              \lambda^2 \delta_2({\bf r-r}') \,]^{-1} \,,
  \end{eqnarray}
These inverse kernels $K({\bf r,r}')$  may be computed by the same
method discussed below  Eq.~(20) if the correct integration area
(or spatial grid) is used as defined in Eqs.~(33) and (37). For the
best choice of the diagonal terms $Q_{ii}$ see Appendix of Ref.~8a.

  The axial magnetic moment of a disk or any other rotationally
symmetric conductor is
  \begin{eqnarray}  % 40
  m(t)=\pi \int_{-b}^b \!\!dy \int_0^{a(y)}\!\! dr\, r^2\,j(r,y) \,.
  \end{eqnarray}
If the applied field is periodic, $H_a(t) = H_0 \sin\omega t$,
the (in general nonlinear) complex susceptibilities of the disk
may be defined as $\chi_\nu = \chi_\nu' -i\chi_\nu''$,
$\nu=1,2,3 \dots$, \cite{8}
  \begin{eqnarray}  % 41
  \chi_\nu(H_0, \omega) = {i \over \pi H_0} \int_0^{2\pi} \!\!
  m(t)\, e^{-i\nu \omega t}\, d(\omega t) \,.
  \end{eqnarray}
Here $\chi_1$ is the fundamental susceptibility and the $\chi_\nu$
with $\nu >1$ correspond to higher harmonics, which are absent
for linear response. The $\chi_\nu$ usually are normalized such that
for $H_0 \to 0$ or $\omega \to \infty$ the ideal diamagnetic
susceptibility $\chi_1(0,\omega)=-1$ results. This is achieved by
dividing all $\chi_\nu$, Eq.~(41), by the absolute value of the
initial slope $[\partial m(H_a) /\partial H_a]_{H_a =0} $.

  As one example, Fig.~12 shows the real and imaginary parts of
the nonlinear fundamental susceptibility
$\chi_1(H_0,\omega)= \chi =\chi'' -i\chi'$ of a thick disk with
$b/a=0.5$ plotted versus the ac amplitude $H_0$ for constant frequency
$\omega = E_c/(\mu_0 j_c a^2)$, creep exponent $n=11$, and for
various London depths $\lambda/a = 0.025 \dots 1$. The same data
are depicted in Fig.~13 as polar plot, $\chi''$ versus $-\chi'$.
In both presentations $\chi(H_0)$ sensitively depends on the
parameters $n$ and $\lambda/a$ and on the geometry. More data
$\chi_\nu (H_0, \omega)$ are available from the author.

\section{Thin plates and films}   % 6

   The geometry of thin plates or planar films with arbitrary shape
in a perpendicular magnetic field $H_a \| z$ differs from the
geometries of Scts.~3 to 5 in that now the current density
is no longer a scalar but has two components,
${\bf j}(x,y,t) = (j_x, j_y)$. However, because of the strict
relation ${\rm div} {\bf j}=0$, this planar ${\bf j}$ may be
derived from a scalar potential (or magnetization, stream function)
$g(x,y,t)$. Since we are interested in the thin film limit we
consider the sheet current
${\bf J}(x,y,t) =(J_x, J_y) = \int {\bf j}(x,y,z,t) \,dz$,
$J_x=\partial g/\partial y$, $J_y = -\partial g/ \partial x$.
The stream lines of ${\bf J}(x,y)$ coincide with the contour
lines $g(x,y)={\rm const}$. Thus, one may choose $g(x,y)=0$
on the edge of the film since the current flows along the
edge. In this section I generalize the theory of
Ref.\ \onlinecite{6} to finite London depth $\lambda$.

   The magnetic field ${\bf H}=H_z(x,y) {\bf \hat z}$ in the film
plane $z=0$ is related to the local magnetization $g(x,y)$ by
an integral over the film area $S$,
  \begin{eqnarray}  % 42
  H_z({\bf r}) =H_a +\int_S\! d^2r'\, Q_{\rm film} ({\bf r,r}')
  \,g({\bf r}')
  \end{eqnarray}
with ${\bf r}=(x,y)$ and  ${\bf r}'=(x',y')$. The integral kernel
  \begin{eqnarray}  % 43
  Q_{\rm film} ({\bf r,r}') =\lim_{z\to 0} {2z^2-\rho^2 \over
  4\pi(z^2+\rho^2)^{5/2} }
  \end{eqnarray}
with $\rho^2 = (x-x')^2+(y-y')^2$, gives the magnetic field
generated by a magnetic dipole of unit strength positioned in the
plane $z=0$ at $(x',y')$. Explicit expressions for films with
rectangular shape are given by Eqs.~(42) to (46) of
Ref.\ \onlinecite{6} in form of Fourier series. To incorporate
a finite London depth $\lambda$ we write the voltage--current
law of the film in the form of Eq.~(5) but now as a function
of the sheet current ${\bf J}(x,y) = {\bf j}(x,y)d(x,y)$. We
consider first a uniform isotropic film with constant thickness
$d$. The local electric field is then
  \begin{eqnarray}  % 44
  {\bf E(J},B) = \rho_s(J,B){\bf J(r},t)+\mu_0 \Lambda
   {\bf \dot J(r},t) \,,
  \end{eqnarray}
where $\rho_s =\rho/d$ is the sheet resistivity and
$\Lambda=\lambda^2/d$ is the effective magnetic penetration depth
of the film. This electric field is related to the induction
${\bf B}=\mu_0{\bf H}$ by the induction law
${\bf \dot B}= -\nabla\times{\bf E}$, which in the film plane
$z=0$ means $\dot B_z = \partial E_x /\partial y -
 \partial E_y/\partial x$. With Eq.~(44) and
$J_x=\partial g/\partial y$, $J_y = -\partial g/ \partial x$,
we obtain
  \begin{eqnarray}  % 45
  \dot B_z({\bf r},t) = \nabla[\, \rho_s \nabla g({\bf r},t)]
     + \mu_0 \Lambda \nabla^2 \dot g({\bf r},t) \,.
  \end{eqnarray}
Taking the time derivative of Eq.~(42), inserting
$\dot B_z = \mu_0 \dot H_z$ from Eq.~(45), and combining the two
terms containing $\dot g$, one arrives at
  \begin{eqnarray}  % 46
  \int_S\! d^2r'\,[\,Q_{\rm film} ({\bf r,r}')-\delta_2({\bf r-r}')
  \Lambda \nabla^2 \,] \, \dot g({\bf r}',t)  \nonumber \\
  = f({\bf r},t) - \dot H_a({\bf r},t) \,,
  \end{eqnarray}
with
  \begin{eqnarray}  % 47
  f({\bf r},t) = \mu_0^{-1} \nabla [\, \rho_s({\bf r},t)
  \nabla g({\bf r},t) ] \,.
  \end{eqnarray}
Inverting this we obtain the equation of motion for $g(x,y,t)$:
  \begin{eqnarray}  % 48
  \dot g({\bf r},t) = \int_S \! d^2r'\, K({\bf r,r}')
  [\, f({\bf r}',t) - \dot H_a({\bf r'},t) ]
  \end{eqnarray}
with $f({\bf r},t)$ from Eq.~(47) and with the inverse kernel
  \begin{eqnarray}  % 49
  K({\bf r,r}')  = [\, Q_{\rm film}({\bf r,r}') -
    \delta_2({\bf r-r}') \Lambda \nabla^2 \,]^{-1} \,.
  \end{eqnarray}
This kernel $K({\bf r}_i, {\bf r}_j)$ may be evaluated on a
grid ${\bf r}_i$ as a Fourier series with discrete $k$ vectors
${\bf K}$. If the Fourier coefficients of
$Q_{\rm film}({\bf r}_i,{\bf r}_j)$ are $Q_{{\bf KK}'}$, given
by Eq.~(46) of Ref.~(6) for a rectangular platelet, then the
Fourier coefficients of $K({\bf r}_i,{\bf r}_j)$ are
the inverse matrix
  \begin{eqnarray}  % 50
   [\, Q_{{\bf KK}'} + \Lambda K^2 \delta_{{\bf KK}'} \,]^{-1} .
  \end{eqnarray}

  Figure 14 shows the current stream lines in a thin rectangle
with side ratio $b/a=0.7$ for two effective penetration depths
$\Lambda/a=0$ and $\Lambda/a =0.1$ and three applied fields
$H_a/J_c = 0.001$, 0.5, and 1.5 (ideal screening, partial
penetration, full penetration). Note that finite $\Lambda $
rounds the sharp corners of the rectangular stream lines and
slightly delays vortex penetration.

   If the superconductor film is {\it nonuniform} [e.g.\ has
spatially varying thickness $d(x,y)$] and/or {\it anisotropic}
[e.g.\ has two (in general nonlinear) resistivities $\rho_{xx}$
and $\rho_{yy}$ and/or two London depths $\lambda_x$ and
$\lambda_y$], then a rather general current--voltage law
${\bf E(J},B,{\bf r}) = (E_x, E_y)$ is
  \begin{eqnarray}  % 51
   E_x=\rho_{sx}J_x +\mu_0 \Lambda_x \dot J_x, ~~
   E_y=\rho_{sy}J_y +\mu_0 \Lambda_y \dot J_y \,,
  \end{eqnarray}
with the two sheet resistivities
  \begin{eqnarray}  % 52
   \rho_{sx}({\bf r})
       ={ \rho_{xx}({\bf J},B,{\bf r}) \over d({\bf r})},~~
   \rho_{sy}({\bf r})
       ={ \rho_{yy}({\bf J},B,{\bf r}) \over d({\bf r})} \,,
  \end{eqnarray}
and two effective penetration depths
  \begin{eqnarray}  % 53
   \Lambda_x({\bf r}) ={ \lambda_x^2({\bf r}) \over d({\bf r})},~~
   \Lambda_y({\bf r}) ={ \lambda_y^2({\bf r}) \over d({\bf r})} \,.
  \end{eqnarray}
The generalized Eqs.~(45) and (46) are then
  \begin{eqnarray}  % 54
   B_z({\bf r},t) &=& {\partial \over\partial x} \Big( \rho_{sy}
   {\partial g \over\partial x} +\mu_0 \Lambda_y
   {\partial \dot g \over\partial x} \Big) \nonumber \\
                  &+& {\partial \over\partial y} \Big( \rho_{sx}
   {\partial g \over\partial y} +\mu_0 \Lambda_x
   {\partial \dot g \over\partial y} \Big) \,,
  \end{eqnarray}
   \vspace{-.4cm} %and
  \begin{eqnarray}  % 55
  \int_S\! d^2r'\, \tilde Q({\bf r,r}') \, \dot g({\bf r}',t)
  = f({\bf r},t) - \dot H_a({\bf r},t) \,,
  \end{eqnarray}
with
  \begin{eqnarray}  % 56, 57
  \tilde Q({\bf r,r}') = Q_{\rm film}({\bf r,r}')
         -\delta_2({\bf r-r}')  \nonumber \\  \times
   \Big( {\partial\over\partial x} \Lambda_y
         {\partial\over\partial x}
       + {\partial\over\partial y} \Lambda_x
         {\partial\over\partial y} \Big)   \,,  \\
  f({\bf r},t) = \mu_0^{-1} \Big[ {\partial\over\partial x}
  \Big(\rho_{sy} {\partial g\over\partial x} \Big)
                                 +{\partial\over\partial y}
  \Big(\rho_{sx} {\partial g\over\partial y} \Big) \Big] \,.
  \end{eqnarray}
The equation of motion for $g(x,y,t)$ now is still given by
Eq.~(48) but with $f({\bf r},t)$ from Eq.~(57) and with the inverse
kernel $K({\bf r,r}') = \tilde Q({\bf r,r}')^{-1}$, Eq.~(56).
The kernels $\tilde Q$ and $K$ may be evaluated by Fourier
transformation.
Some figures of current stream lines and field profiles
for thin rectangles with anisotropic
pinning and $\Lambda=0$ are depicted in Ref.\ \onlinecite{28}.

   When $\Lambda=0$, the inverse kernel $K=Q_{\rm film}^{-1}$
can be obtained by iteration, \cite{29} see also
the Appendix of Ref.\ \onlinecite{27}:
  \begin{eqnarray}  % 58
 K({\bf r, r}') =\! {1 \over C({\bf r})} \Big[ \delta({\bf r\!-\!r}')
  \!+\!\!\! \int_S \!\! { K({\bf r}'',{\bf r}')\! -\!K({\bf r, r}')
  \over  4\pi |{\bf r- r}''|^3 } d^2r'' \Big] \!\!\!
%  C({\bf r}) K({\bf r, r}') =  \delta({\bf r,r}')
%  + \int_S {  K({\bf r}'',{\bf r}) -K({\bf r, r}') \over
%  4\pi |{\bf r- r}'|^3 }  \,,
  \end{eqnarray}
starting with $ K({\bf r, r}') = 0$. Here $C({\bf r})$ is defined
as an integral over the infinite area $\bar S$ outside the film
or as a contour integral along the film edge:
  \begin{eqnarray}  % 59
   4 \pi C({\bf r}) = \int_{\bar S}  { d^2 r' \over R^3} \,=
      \int_0^{2\pi} \! {d\phi \over R(\phi) }
  \end{eqnarray}
with $R=|{\bf r-r}'|$. For rectangular films ($|x| \le a$,
$|y| \le b$)  one has explicitly
  \begin{eqnarray}  % 60
  C(x,y) ={1\over 4\pi } \sum_{p,q} \Big[ (a-px)^{-2}
                                         +(b-qy)^{-2} \Big]
  \end{eqnarray}
with $p,q=\pm 1$. Expression (60) may also be used for rectangular
films with holes or slots if the integral in Eq.~(58) is restricted
to the area of the real film.

  For numerics one needs the discretized versions of Eq.~(42) and
its inverse. Introducing a 2D grid ${\bf r}_i = (x_i, y_i)$ with
weights $w_i$ [\,$\sum_i w_i = S $, cf.\ Eq.~(21)] and writing
$Q_{ij} = Q({\bf r}_i,{\bf r}_j)$, $K_{ij} = K({\bf r}_i,{\bf r}_j)$,
$h_i = H_z({\bf r}_i) -H_a$, and $g_i = g({\bf r}_i)$ one has
  \begin{eqnarray}  % 61
  h_i = \sum_j Q_{ij} w_j g_j ,~~~ g_i = \sum_j K_{ij} h_j \,.
  \end{eqnarray}
The inverse kernel $K_{ij} = (Q_{ij}w_j)^{-1}$ cannot be obtained
directly by discretizing $Q_{\rm film}({\bf r,r}')$, Eq.~(43), but
it may be obtained by iterating Eq.~(58). This iteration can be
solved in one step by inverting a matrix,
  \begin{eqnarray}  % 62
  K_{ij} = \Big[ \delta_{ij} \Big( C_i +\sum_{l\ne i} w_l q_{il}
  \Big)  -(1-\delta_{ij}) w_j q_{ij} \Big]^{-1} ,
  \end{eqnarray}
where $C_i =C({\bf r}_i)$ and
$q_{ij} = 1/(4\pi |{\bf r}_i-{\bf r}_j|^3)$. Note that the terms
in Eq.~(62) must not be rearranged since $q_{ii} = \infty$.

  Using this matrix $K_{ij}$ the current stream lines in
ideally screening flat films of arbitrary shape or of films with
trapped vortices are easily calculated. If the applied field is
constant, $h_i = -H_a$, the stream lines are the contour lines of
the trace of matrix $K_{ij}$:
  \begin{eqnarray}  % 63
  g_i = -H_a \sum_j K_{ij} \,.
  \end{eqnarray}
If a vortex with flux $\Phi_0$ is trapped at a grid point
${\bf r}_j$ in the film, the magnetic field is $H_z({\bf r}) =
\Phi_0 \delta({\bf r-r}_j) \approx (\Phi_0/w_j)\delta_{ij}$ at
${\bf r = r}_i$, since we assume here $\Lambda=0$.
Thus, $g_i$ is just one row of the matrix:
  \begin{eqnarray}  % 64
  g_i = (\Phi_0 / w_j) \cdot K_{ij} \,.
  \end{eqnarray}
In general, the stream function $g_i$ is the linear superposition
of the contributions of the applied field and trapped vortices,
and possibly of currents applied by contacts.

  The following relations are useful:
The difference $g_i -g_j$ equals the total current crossing any line
connecting the points ${\bf r}_i$ and ${\bf r}_j$; the function
 $\pm g({\bf r})$ is the potential in which a probe vortex
(or flux bundle) of appropriate sign moves;
the Lorentz force on a vortex is perpendicular to the contour
lines of $g({\bf r})$; and the matrix $K_{ij}$ is proportional to
the interaction energy between two vortices at ${\bf r}_i$ and
${\bf r}_j$. All these statements apply to any film shape and to
any $\Lambda$ if the general kernel $K$, Eq.~(49), is used.
For $\Lambda=0$, the kernel $K$ is given by Eqs.~(58) or (62).

Thus, a vortex will nucleate at the position on the film edge where
the stream lines are densest, and then move to an extremum of
$g({\bf r})$ if it is not pinned by material inhomogeneities.

  Figure 15 shows the so called washer geometry that is used for
SQUIDs (Superconducting Quantum Interference Devices), \cite{30}
namely, a superconductor thin square or rectangle (or circular disk)
with a slot and central hole. See Refs.~\onlinecite{31,32,33,34}
for theories of such washers with finite effective penetration
depth $\Lambda$. After nucleation at one border of the slot, a
vortex without pinning will move towards a maximum of $g({\bf r})$.
For the square washer in Fig.~15 (top left) there is one flat
maximum of $g({\bf r})$ to the left of the central hole. For the
rectangular washer (bottom) there are three such maxima, to the
left of the center and above and below the slot, which are
separated by two saddle points above and below the central hole.
A vortex may thus enter near the middle of the slot,
jump to the nearby maximum of $g$, and then jump over the
saddle point to the wide maximum at the left. The field-driven
thermally activated motion of vortices generates low-frequency
noise in the SQUID, which may be suppressed by introducing pinning
centers, e.g.\ tiny holes. \cite{35}

 In the examples of Fig.~15, $\Lambda = 0$ (or $\Lambda \ll a$)
was assumed. In this case the vortices interact with the screening
currents of the applied field and with other vortices only via the
stray field outside the film. In general, at distances
$r \ll \Lambda$ the point vortices in thin films interact via a
logarithmic potential similar to the interaction of line vortices
in the bulk. At distances $r \gg \Lambda$ the interaction is
mediated by the stray field, is of long range, and depends on
the shape of the film. This long range part of the vortex
interaction (or flux-bundle interaction) is given by the
integral kernel $K$, Eq.~(58) and (62). For infinitely extended
films the $\Lambda$ dependent vortex interaction and the
magnetic field of a vortex in thin films ($d \ll \lambda$) were
derived by Pearl, \cite{12} see also Refs.\ \onlinecite{2,36}.
For infinite films of arbitrary thickness the vortex interaction
and magnetic field are obtained in Refs.~\onlinecite{37,38}.

\section{Summary}   % 7

  In this paper finite London penetration depth $\lambda$ is
introduced into known continuum methods which
compute the electromagnetic response of type-II superconductors
in various geometries. In addition, solution methods for some
new geometries are presented, namely, thick and thin strips with
transport current and with both applied current and applied
magnetic field, and thin films of arbitrary shape, e.g., the
washer geometry used for SQUIDs. The inverse integral kernel
$K({\bf r,r}')$ or the matrix $K_{ij}$ defined in Eqs.~(58) and
(62) have the simple physical interpretation of the interaction
energy between two point vortices in a film of given shape.
The stream function $g({\bf r})$ of Eqs.~(63) and (64) is just
the potential in which a probe vortex (or flux bundle) in the
film moves and which is caused by the applied
perpendicular magnetic field (and/or applied current)
and by the other vortices. In the depicted case of
small $\Lambda= \lambda^2 /d$, this interaction is
mediated only via the magnetic stray field outside the thin film,
which crucially depends on the shape of the film. For finite films
with arbitrary $\Lambda$, the vortex interaction is
implicitly given by the inverse kernel $K({\bf r,r}')$, Eq.~(49).
Explicit $\Lambda$ dependent expressions for concrete film shapes
will be given elsewhere.

\acknowledgements

  The author wishes to acknowledge the hospitality of the
Institute of Electronic and Superconducting Materials,
University of Wollongong, Australia, where part of this work was
performed, and financial support from the Australian Research
Council, IREX Programm.

%xxxxxxxxxxxxxxxxxxxxxxxxxxxxxxxxxxxxxxx
\vspace{-0.4 cm} %   \vspace{-0.3 cm}  %-.6

% \end{multicols}
% \end{document}
 \vspace{0.2cm}

 \begin{figure}[F1]
\epsfxsize= .85\hsize  \vskip 1.0\baselineskip
\centerline{ \epsffile{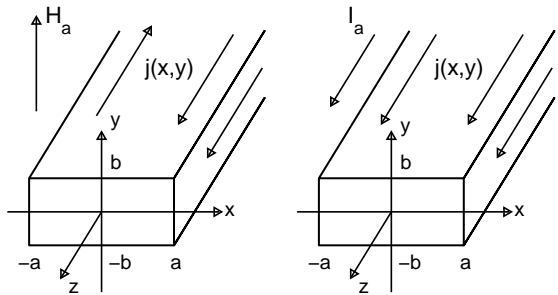}}
 \vspace{.3cm}
\caption{Geometry of long superconductor strips with rectangular
cross section $2a \times 2b$ in a perpendicular applied magnetic
field $H_a$ (left) or with applied electric current
$I_a = I =\int j(x,y) \,dx\,dy$ (right).
 } \end{figure}
 \vspace{-.3cm}

 \begin{figure}[F2]
\epsfxsize= .98\hsize  \vskip 1.0\baselineskip
\centerline{ \epsffile{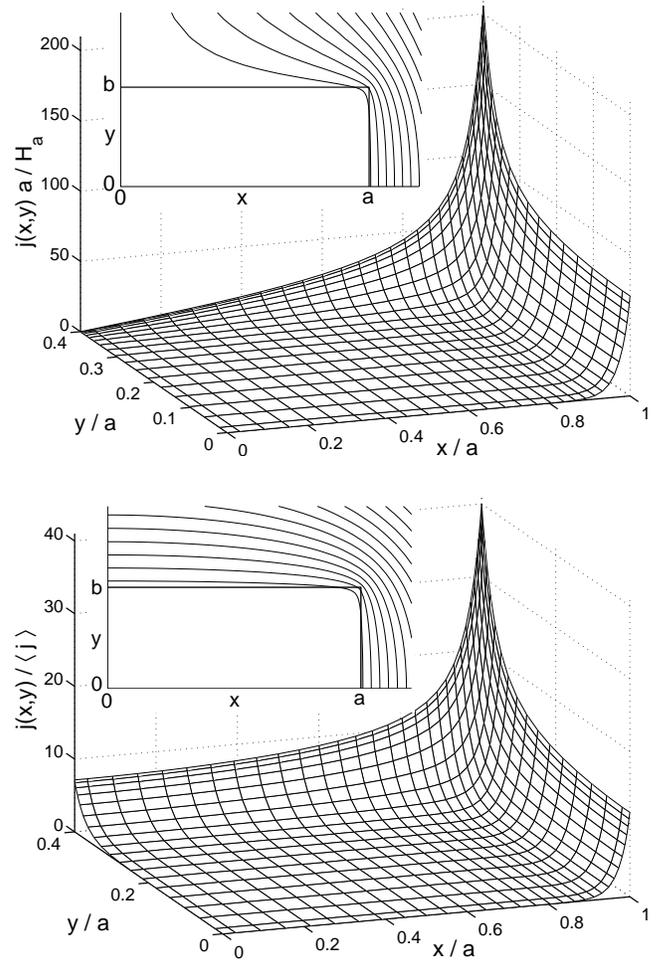}}
 \vspace{.3cm}
\caption{A strip with aspect ratio $b/a=0.4$,
cf.~Fig.~1, in the Meissner state with London penetration depth
$\lambda = 0.025 a$.  Shown is the current density in a quarter
of the cross section and the magnetic field lines (inset).
Top: In perpendicular applied magnetic field $H_a$.
Bottom: With applied current.
 } \end{figure}
  % .  \vspace{27cm} . \newpage

 \begin{figure}[F3]
\epsfxsize= .98\hsize  \vskip 1.0\baselineskip
\centerline{ \epsffile{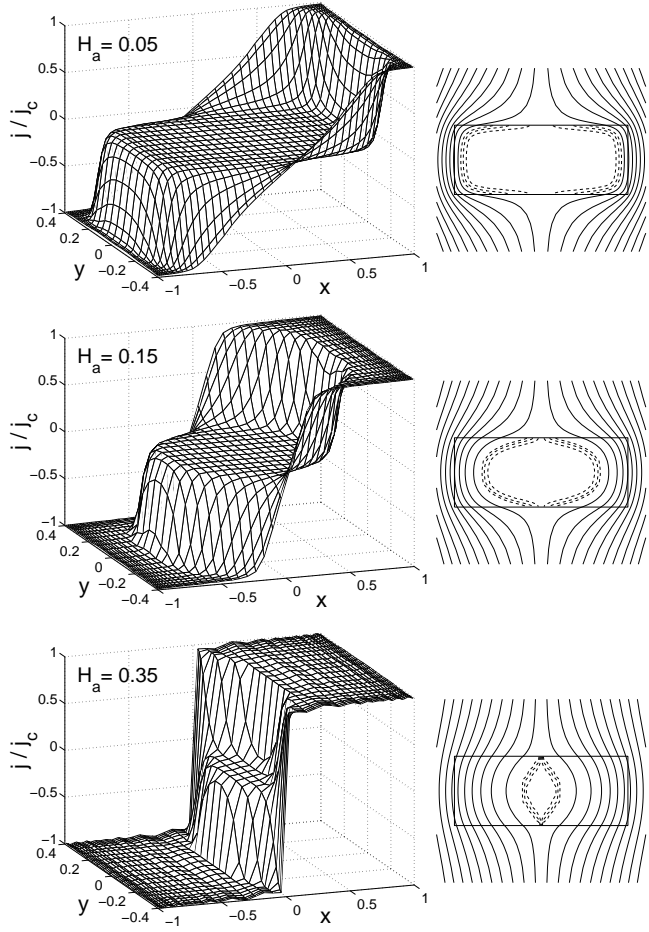}}
 \vspace{.3cm}
\caption{Current density $j(x,y)$ (left) and magnetic field lines
(right) of a superconductor strip with aspect ratio $b/a=0.4$ in
increasing magnetic field $H_a = 0.05$, 0.15, and 0.35 in units
$ a j_c$ ($j_c$ = critical current density). The dashed lines
are contours of the current density at $j/j_c=\pm 0.75$, $\pm 0.45$,
and $\pm 0.15$. The superconductor is characterized by a
pinning-caused voltage--current law $E \propto (j/j_c)^{101}$ and
by a small London penetration depth $\lambda / a = 0.025$.
  } \end{figure}
  % \vspace{-.4cm}

 \begin{figure}[F4]
\epsfxsize= .98\hsize  \vskip 1.0\baselineskip
\centerline{ \epsffile{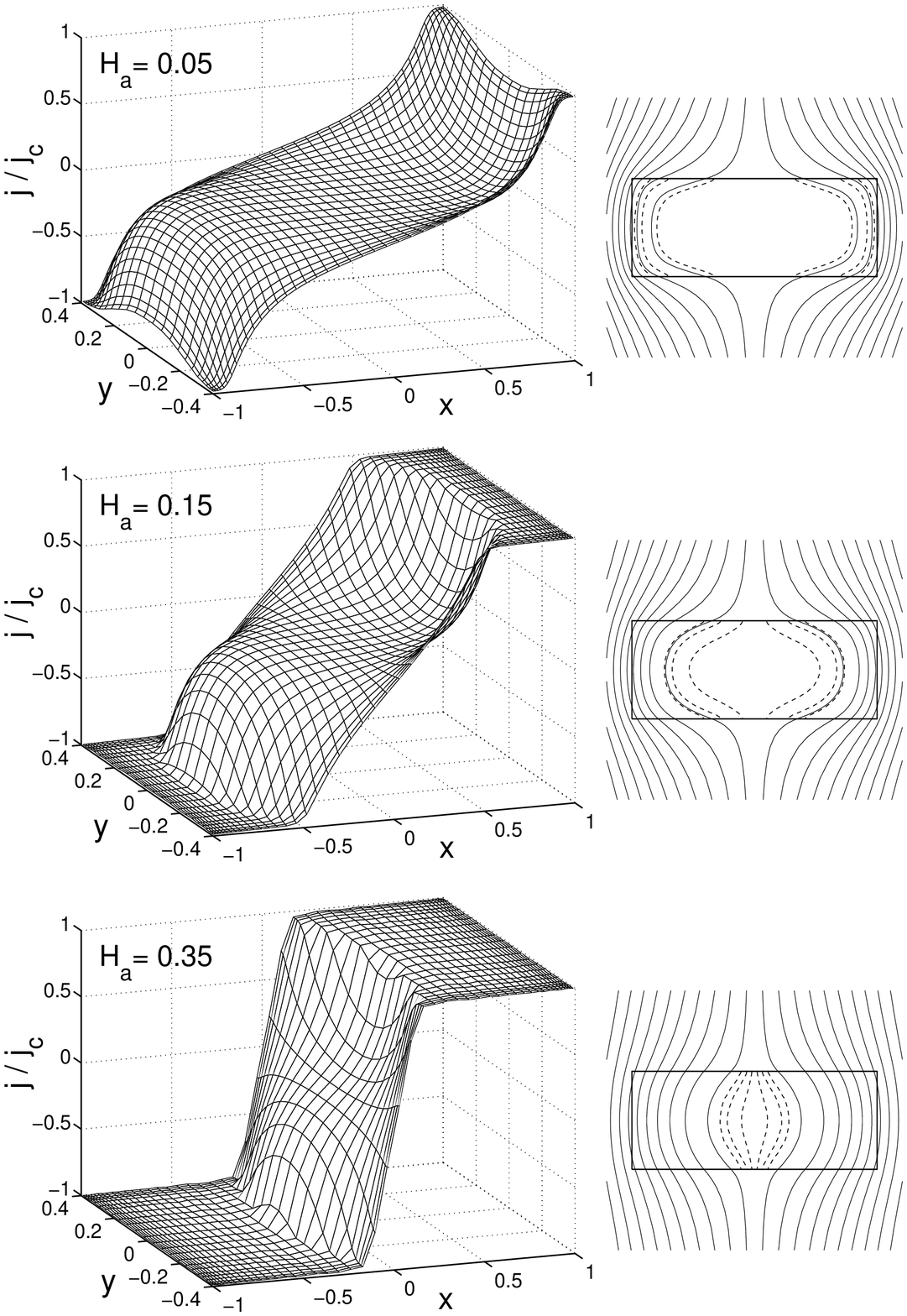}}
 \vspace{.3cm}
\caption{As Fig.~3 but for larger London depth $\lambda/a = 0.1$.
  } \end{figure}

 \begin{figure}[F5]
\epsfxsize= .98\hsize  \vskip 1.0\baselineskip
\centerline{ \epsffile{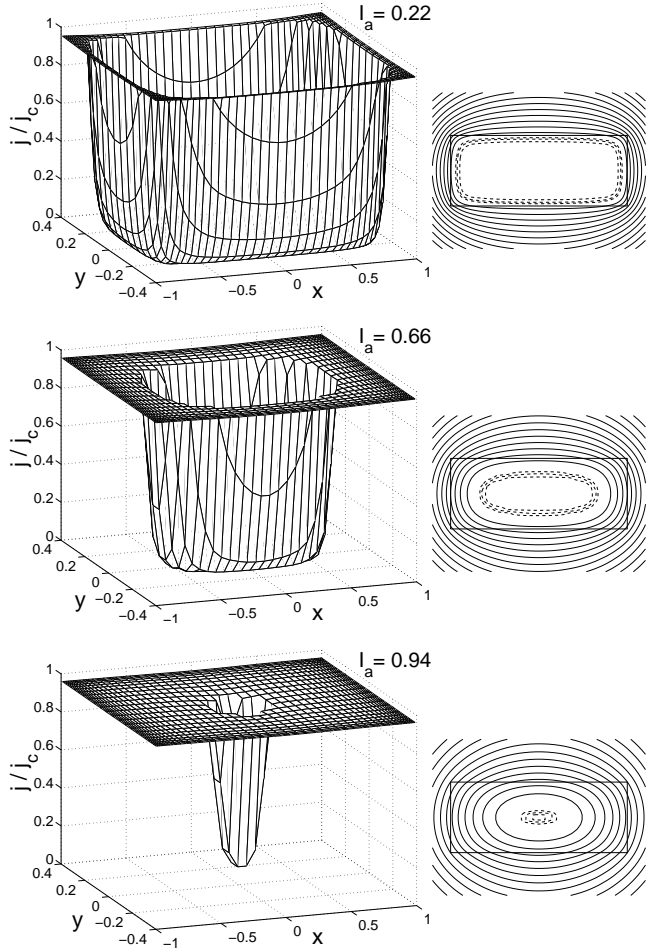}}
 \vspace{.1cm}
\caption{Current density $j(x,y)$ (left) and magnetic field lines
(right) of a superconductor strip with aspect ratio $b/a=0.4$ with
applied current $I_a = 0.22$, 0.66, and 0.94 in units of the
critical current $I_c = 4abj_c$.  The dashed lines
are contours of the current density at $j/j_c= 0.75$, 0.45, and
 0.15. As in Figs. 3, 4 a voltage--current law
$E \propto (j/j_c)^{101}$ was used and a small London penetration
depth $\lambda / a = 0.025$.
  } \end{figure}

 \begin{figure}[F6]
\epsfxsize= .97\hsize  \vskip 1.0\baselineskip
\centerline{ \epsffile{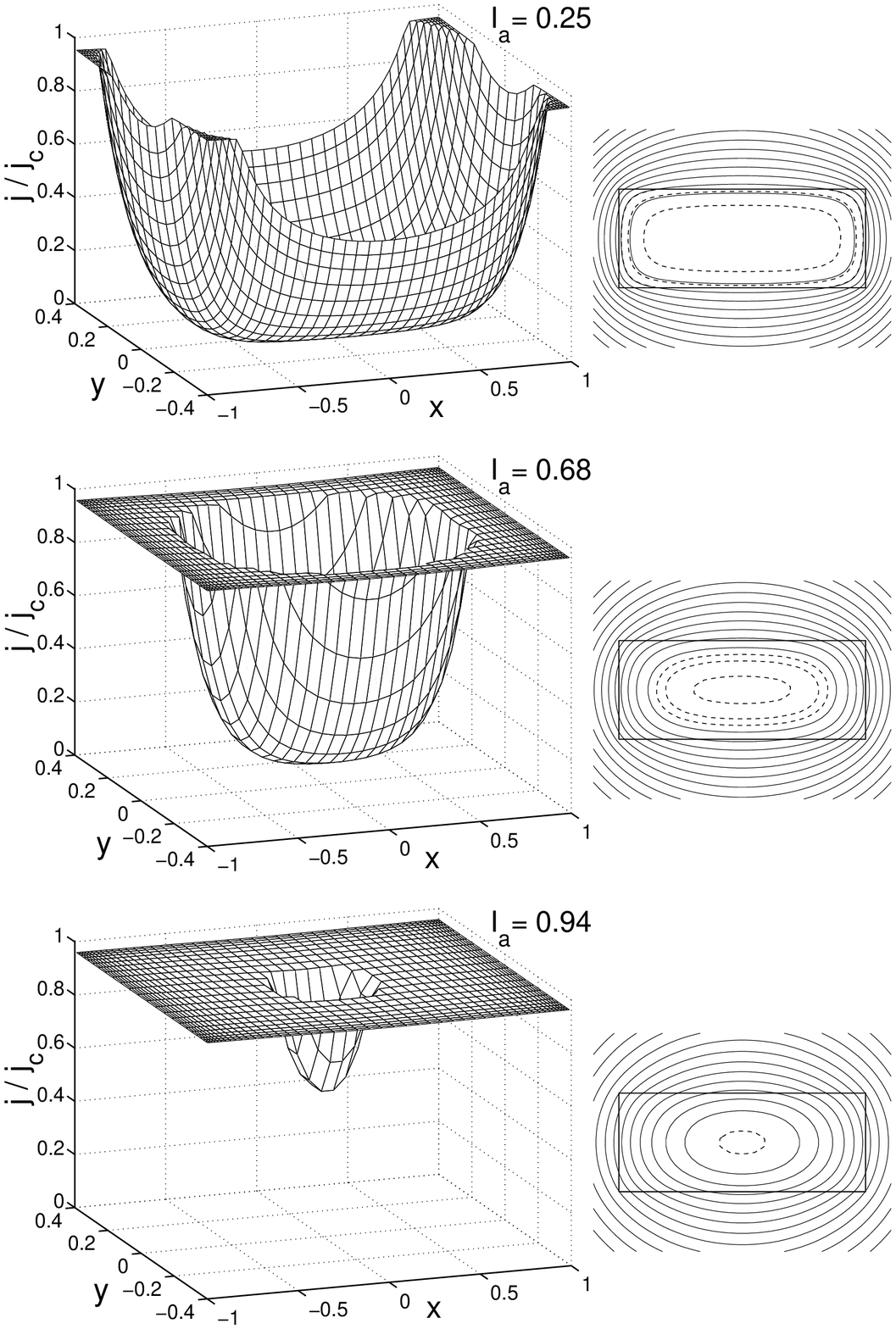}}
 \vspace{.1cm}
\caption{As Fig.~5 but for larger London depth $\lambda/a = 0.1$.
  } \end{figure}
 \vspace{-1.cm} \newpage

 \begin{figure}[F7]
\epsfxsize= .98\hsize  \vskip 1.0\baselineskip
\centerline{ \epsffile{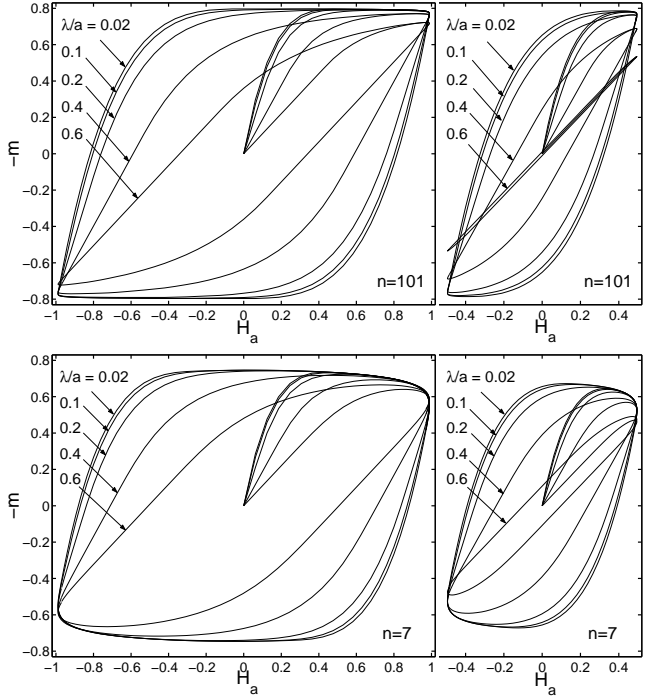}}
 \vspace{.2cm}
\caption{Magnetization curves $m(H_a)$ of a superconductor thick
strip with aspect ratio $b/a=0.4$ in a perpendicular field
$H_a(t)= H_0 \sin{\omega t}$ at two amplitudes $H_0/H_p = 2$ (left)
and $H_0/H_p = 1$ (right) with $H_p = 0.4945 aj_c $ the field of
full penetration. Fields in units $a j_c$ and magnetic moment
$m$ in units $a^3 j_c$ (per unit length).
Shown are virgin curves and hysteresis loops for two creep
exponents $n=101$ (top) and $n=7$ (bottom) and for London
depths $\lambda/a = 0.02$, 0.1, 0.2, 0.4, and 0.6.
  } \end{figure}

 \begin{figure}[F8]
\epsfxsize= .98\hsize  \vskip 1.0\baselineskip
\centerline{ \epsffile{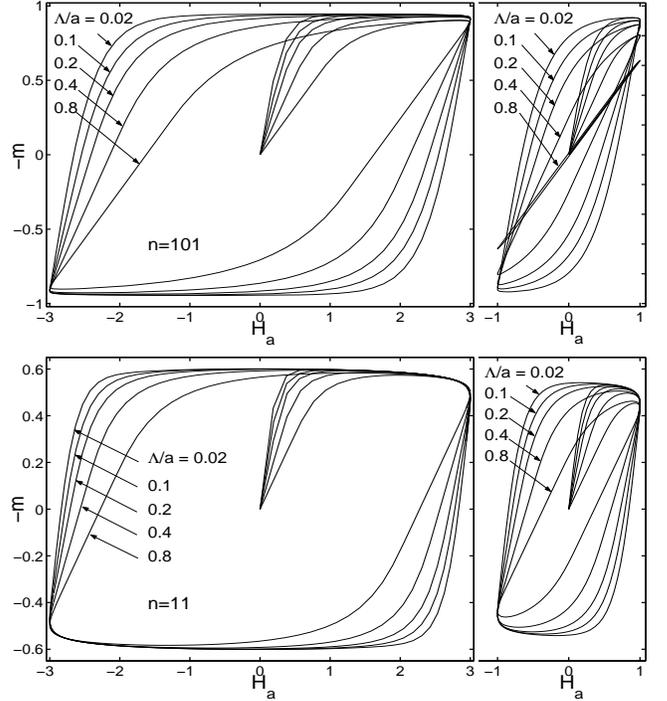}}
 \vspace{.2cm}
\caption{Magnetization curves $m(H_a)$ of a superconductor thin
strip with aspect ratio $b/a=0.001$ in a perpendicular field
$H_a(t)= H_0 \sin{\omega t}$ at two amplitudes $H_0 = 3$ (left)
and $H_0 = 1$ (right) in units of the critical sheet current
$J_c =d j_c$ ($d=2b$). Magnetic moment $m$ in units
$a^2 J_c$ (per unit length), $H_p=2.52 J_c$. Virgin curves and
hysteresis loops for two creep exponents $n=101$ (top) and $n=11$
(bottom) and five effective penetration depths
$\Lambda = \lambda^2/d$:
$\Lambda/a = 0.02$, 0.1, 0.2, 0.4, and 0.8.
  } \end{figure}

 \begin{figure}[F9]
\epsfxsize= .98\hsize  \vskip 1.0\baselineskip
\centerline{ \epsffile{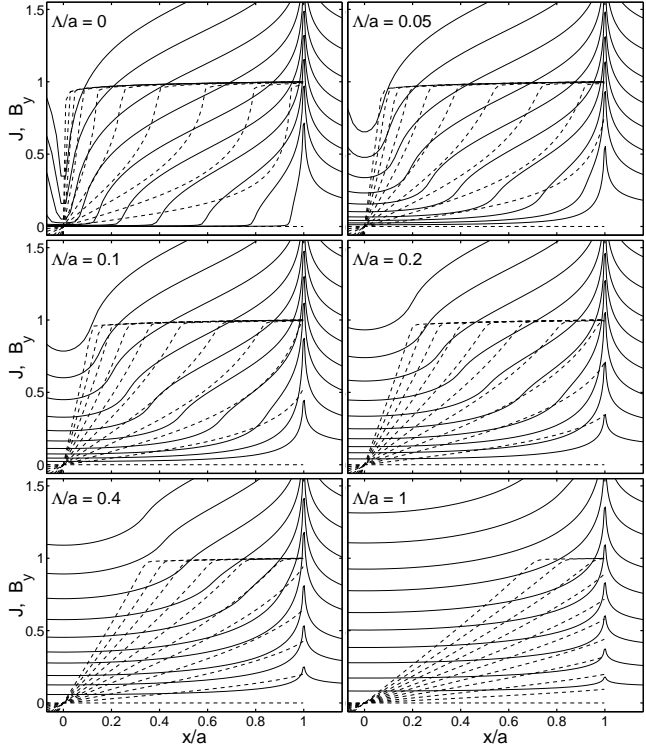}}
 \vspace{.2cm}
\caption{Profiles of sheet current $J(x)$ (dashed lines) and
perpendicular induction $B_y(x)$ (solid lines) of a thin strip in
increasing perpendicular magnetic field
$H_a=0.10$, 0.22, 0.35, 0.49, 0.65, 0.82, 1.0, 1.23, 1.46, and
1.72 (curves from bottom to top). $J$ and $H_a$ are in
units $J_c=d j_c$ ($d=2b \ll a$), $B$ in units $\mu_0 J_c$.
For six effective penetration depths $\Lambda=\lambda^2/d$:
$\Lambda/a = 0$, 0.05, 0.1, 0.2, 0.4, and 1. Creep exponent $n=51$.
  } \end{figure}

 \begin{figure}[F10]
\epsfxsize= .98\hsize  \vskip 1.0\baselineskip
\centerline{ \epsffile{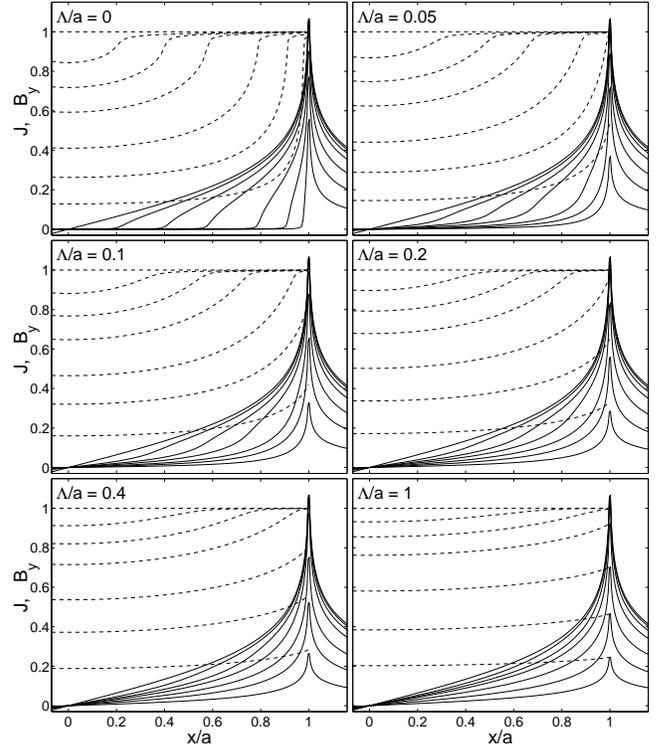}}
 \vspace{.2cm}
\caption{As Fig.~9 but for strips with increasing applied current
$I_a/I_c= 0$, 0.2, 0.4, 0.6, 0.8, 0.9, 0.965, and 1.
$I_c = 4ab j_c$ is the critical current of the strip.
  } \end{figure}
  \vspace{-.7cm}

 \begin{figure}[F11]
\epsfxsize= .80\hsize  \vskip 1.0\baselineskip
\centerline{ \epsffile{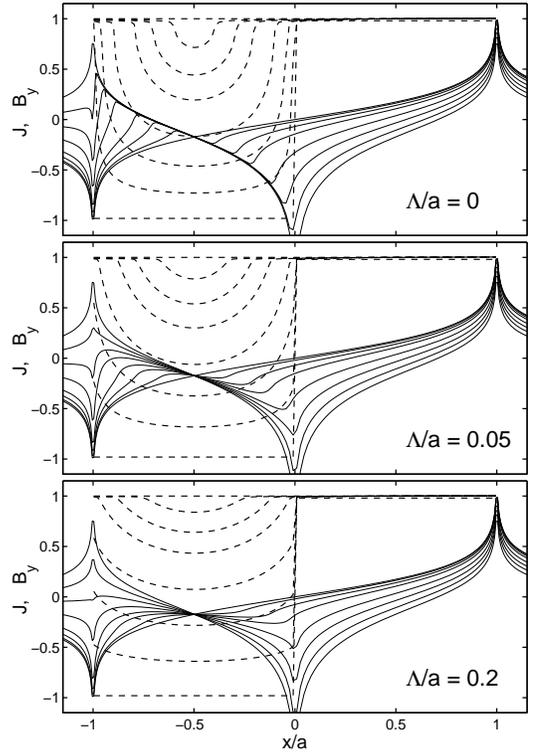}}
 \vspace{.2cm}
\caption{As Fig.~9 but for a thin strip in large
magnetic field $H_a \gg H_p$ to which an increasing current is
applied, $I_a/I_c = 0$, 0.2, 0.4, 0.6, 0.8, 0.9, 0.97, and 1
(from bottom to top) with $I_c = 4ab j_c$.
$\Lambda/a = 0$, 0.05, and 0.2.  $n=51$.
  } \end{figure}

 \begin{figure}[F12]
\epsfxsize= .98\hsize  \vskip 1.0\baselineskip
\centerline{ \epsffile{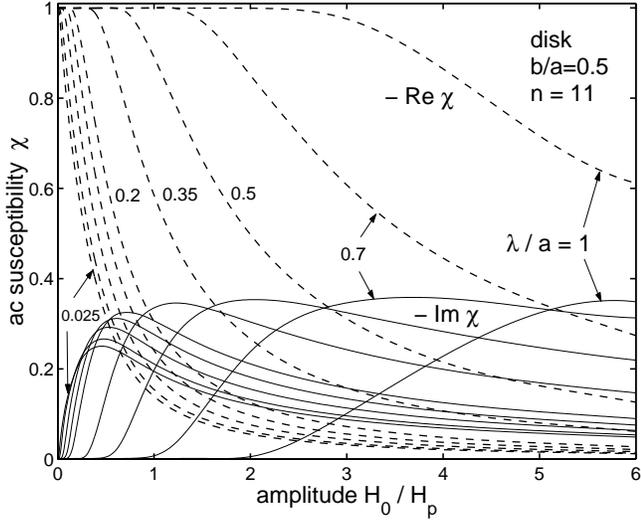}}
 \vspace{.2cm}
\caption{The nonlinear complex ac susceptibility
$\chi(H_0,\omega) = \chi' - i\chi''$
of a type-II superconductor thick disk with radius $a$ and
half thickness $b=0.5 a$ plotted versus the amplitude $H_0$
of the ac magnetic field in units of the Bean field of full
penetration $H_p = 0.722 a j_c$. The curves are for
different London depths $\lambda/a = 0.025$, 0.05, 0.1, 0.15,
0.2, 0.35, 0.5, 0.7, and 1. $E_v \propto j^n$, $n=11$,
$\omega = E_c/(\mu_0 j_c a^2)$.
  } \end{figure}

 \begin{figure}[F13]
\epsfxsize= .98\hsize  \vskip 1.0\baselineskip
\centerline{ \epsffile{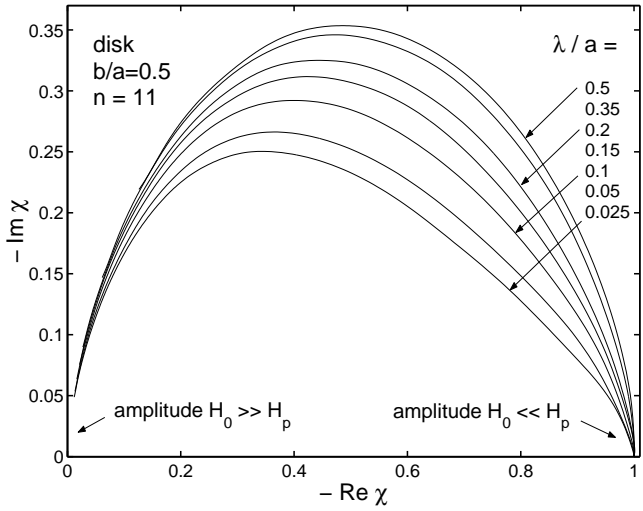}}
 \vspace{.2cm}
\caption{The nonlinear susceptibility of Fig.~12 plotted
as $\chi''$ versus $-\chi'$.
  } \end{figure}
  \vspace{-.5cm}

 \begin{figure}[F14]
\epsfxsize= .9\hsize  \vskip 1.0\baselineskip
\centerline{ \epsffile{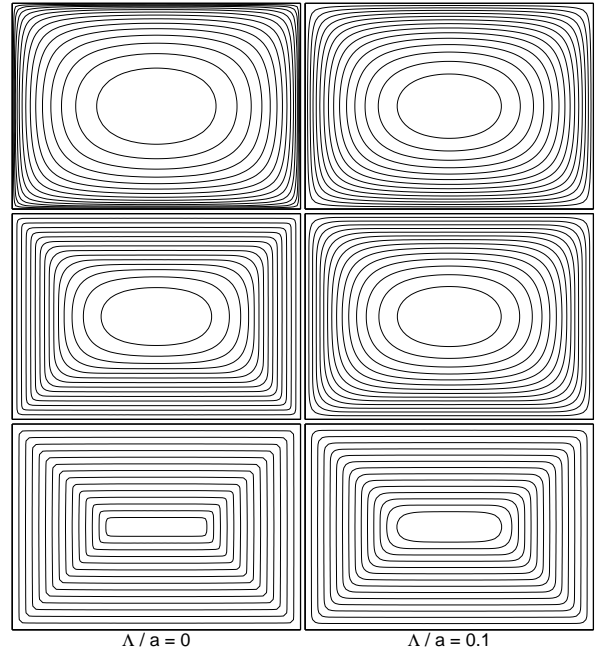}}
 \vspace{.2cm}
\caption{The current stream lines in a thin superconductor
rectangle ($b/a=0.7$)
with $\Lambda/a =0$ (left) and $\Lambda/a = 0.1$ (right) at applied
perpendicular fields (from top to bottom) $H_a/J_c = 0.001$,
 0.5, and 1.5.
  } \end{figure}
  \vspace{-.6cm}

 \begin{figure}[F15]
\epsfxsize= .9\hsize  \vskip 1.0\baselineskip
\centerline{ \epsffile{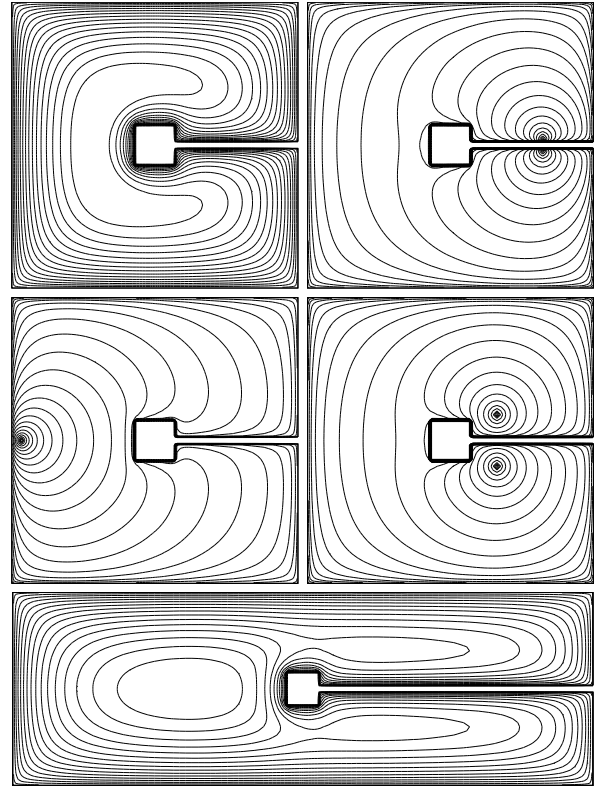}}
 \vspace{.2cm}
\caption{Thin superconductor square ($a=b$) and rectangle ($a/b=3$)
with slit and hole (washer) in the Meissner state with $\Lambda=0$.
Shown are the current stream lines for constant applied field
(top left, bottom) and for zero applied field but with one or two
vortices trapped at various positions.
  } \end{figure}

\end{multicols}
\end{document}